# Direct demonstration of topological stability of magnetic skyrmions via topology manipulation

Soong-Geun Je,[1,2,3,4] Hee-Sung Han,[5] Se Kwon Kim,[6] Sergio A. Montoya,[7] Weilun Chao,[1] Ik-Sun Hong,[8] Eric E. Fullerton,[9,10] Ki-Suk Lee,[5] Kyung-Jin Lee,[8,11] Mi-Young Im,[1,2,5,†] and Jung-Il Hong[2,†]

[1]Center for X-ray Optics, Lawrence Berkeley National Laboratory, Berkeley, CA 94720, USA.

[2]Department of Emerging Materials Science, DGIST, Daegu 42988, Korea.

[3]Center for Spin-Orbitronic Materials, Korea University, Seoul 02841, Korea.

[4]Department of Physics, Chonnam National University, Gwangju 61186, Korea

[5]School of Materials Science and Engineering, Ulsan National Institute of Science and Technology, Ulsan 44919, Korea.

[6]Department of Physics and Astronomy, University of Missouri, Columbia, MO 65211, USA.

[7]Space and Naval Warfare Systems Center Pacific, San Diego, CA 92152, USA.

[8]KU-KIST Graduate School of Converging Science and Technology, Korea University, Seoul 02841, Korea.

[9]Center for Memory and Recording Research, University of California–San Diego, La Jolla, CA 92093, USA.

[10]Department of Electrical and Computer Engineering, University of California–San Diego, La Jolla, CA 92093, USA.

[11]Department of Materials Science and Engineering, Korea University, Seoul 02841, Korea.

†e-mail: mim@lbl.gov; jihong@dgist.ac.kr

**Abstract**

Topological protection precludes a continuous deformation between topologically inequivalent configurations in a continuum. Motivated by this concept, magnetic skyrmions, topologically nontrivial spin textures, are expected to exhibit the topological stability, thereby offering a prospect as a nanometer-scale non-volatile information carrier. In real materials, however, atomic spins are configured as not continuous but discrete distribution, which raises a fundamental question if the topological stability is indeed preserved for real magnetic skyrmions. Answering this question necessitates a direct comparison between topologically nontrivial and trivial spin textures, but the direct comparison in one sample under the same magnetic fields has been challenging. Here we report how to selectively achieve either a skyrmion state or a topologically trivial bubble state in a single specimen and thereby show how robust the skyrmion structure is in comparison with the bubbles for the first time. We demonstrate that topologically nontrivial magnetic skyrmions show longer lifetimes than trivial bubble structures, evidencing the topological stability in a real discrete system. Our work corroborates the physical importance of the topology in the magnetic materials, which has hitherto been suggested by mathematical arguments, providing an important step towards ever-dense and more-stable magnetic devices.

I. INTRODUCTION

Topology [1,2] constitutes profound viewpoints in modern physics on revealing various robust states existing in many physical systems including topological insulators [3,4], ultracold atoms [5], topological insulator lasers [6] and topological mechanical metamaterials [7]. Topology has also been remarkably successful in representing various magnetic phenomena [1,8-18]. In a continuum description of magnetic systems, the topological protection means that a continuous deformation between spin structures with different topologies is not allowed [1,2]. This implies exceptional stability of spin textures with nontrivial topology against collapse to a uniform spin configuration.

A compelling example of such topological spin structures is the magnetic skyrmion. It is a swirling spin structure [Fig. 1(a)] possessing a quantized topological charge defined by $Q = \frac{1}{4\pi}\int \bm{m} \cdot (\partial_x \bm{m} \times \partial_y \bm{m}) dx dy = \pm 1$, which measures how many times $\bm{m}$ winds the unit sphere within a closed surface, where $\bm{m}$ is the unit vector of magnetization. In terms of topological charge, the skyrmion structure is topologically distinguished from a uniform ferromagnetic state with $Q = 0$, hence characterized as a topologically nontrivial spin texture.

The magnetic skyrmion with exceptionally small size is anticipated to remain robust and can be driven by electric currents easily without being interrupted by system disorders such as structural defects [19-28]. Therefore, it is being considered as a promising candidate for information carrier in the applications for ultrahigh density data storage [19,20,29], logic [30], and neuromorphic [31,32] technologies. As the successful implementation of such skyrmion-based applications crucially relies on the retention of skyrmion structures, the topologically protected property of the skyrmions is the most important prerequisite.

In real materials, however, magnetic moments are localized on atoms in a discrete lattice so that the collapse of magnetic skyrmions is possible by overcoming a finite energy barrier in an atomistic length scale, where the topological argument is nullified. This realistic situation thus naturally raises fundamentally and technologically important questions: Is the topological protection still a viable concept to guarantee the skyrmion stability and how strong is the constraint in the real discrete system?

A lot of efforts have been devoted to addressing the issues on the skyrmion stability in a real system, but they still remain controversial [33]. Theoretical works have suggested that there are alternative skyrmion decay paths along which the energy barrier is lower than that of atomistic-shrinking of skyrmions [34-37] or the entropy has an important role in the skyrmion lifetime [37-39]. Several experiments [40-43] have been carried out to demonstrate the topological protection, but their critical limitation is the absence of experimental comparison with a topologically trivial counterpart of the skyrmion, which is necessary to address the role of the topological nature of a skyrmion.

A representative of the topologically trivial counterpart to the skyrmion is a bubble of $Q = 0$ as schematically depicted in Fig. 1(b). The boundary of a bubble forms with a pair of half circles with winding spins in the opposite direction, and two half circles are connected by Bloch lines. Since its topological charge $Q$ is 0, the decay of the bubble into a uniformly magnetized state corresponds to the case where the topological effect does not participate.

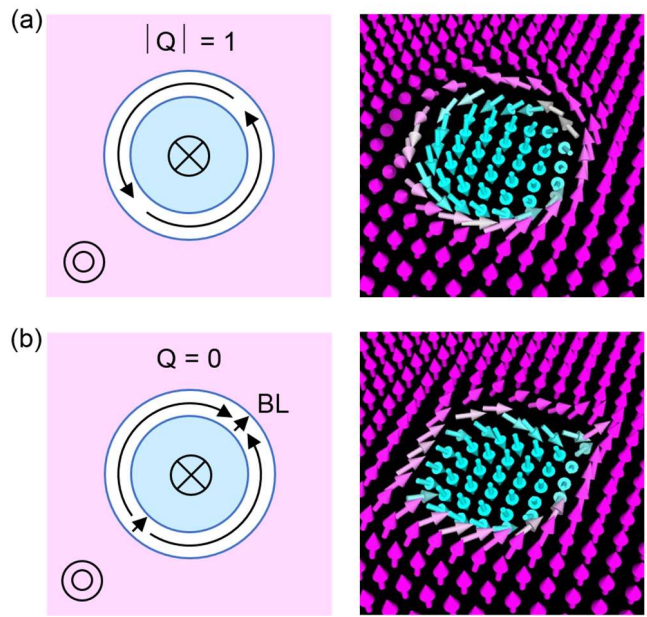

**FIG. 1.** *Schematic illustration of a topologically nontrivial skyrmion and a topologically trivial bubble.* (a) Bloch-type skyrmion with $|Q| = 1$. (b) Bubble with $Q = 0$. The BL refers to the Bloch line.

The experimental difficulties hindering a thorough assessment of the topological effect to the skyrmion stability is twofold. Firstly, experimental measurement of sub-100 nm spin structures with a sufficiently high temporal resolution to detect the skyrmion lifetime is technically a challenging issue. Secondly and more importantly, tailoring topology [42-44], thereby selectively preparing either skyrmions or bubbles in the same material, is another challenging issue which has not been fully explored. A measurable physical quantity to estimate the topological effect is the lifetime, which however varies drastically depending on material properties. Therefore, resolving the second issue is particularly important to directly compare the stabilities of topologically nontrivial and trivial structures on an equal footing.

II. STABILITY OF TOPOLOGICAL SPIN STRUCTURES

Here, by overcoming these two challenges, we experimentally demonstrate the stability of magnetic skyrmions rooted in topology. By choosing different magnetic field pathways, we are able to selectively reach either a magnetic skyrmion state or a bubble state in the same specimen under the same magnetic field strength, providing a versatile route towards the topology manipulation and fair comparison of the topological effect. The lifetimes of both skyrmions and bubbles are then studied using the full-field transmission soft x-ray microscopy (MTXM) combined with pulsed magnetic fields of various durations to enhance the time resolution. We find that magnetic skyrmions exhibit much longer lifetime than bubbles even at higher fields. This result is in line with the expectation for the topological protection: the topology change is involved in the annihilation of skyrmions whereas it is not in the annihilation of bubbles. This work provides the first experimental evidence for the existence of the topological stability of skyrmions in a real discrete system.

A. Selective generation of magnetic bubbles and skyrmions

For this study, the material we selected is an amorphous ferrimagnetic Fe/Gd multilayer film [Fe(0.34 nm)/Gd(0.4 nm)]×120. In a previous work [45], Lorentz transmission electron microscopy and MTXM study confirmed that this material exhibits a dipolar-stabilized skyrmion lattice upon increasing out-of-plane (OOP) fields by pinching off stripe domains. Owing to the lack of the Dzyaloshinskii-Moriya interaction, the skyrmions in this material are of Bloch type [45] at the middle layer of the film as shown in Fig. 1(a) (see Supplementary Material 1 for the generation process and Supplementary Material 2 for the structure of the skyrmion). Also, topologically trivial bubbles coexisting with skyrmions are often observed [46], suggesting that this material is able to serve as an ideal testbed for the comparison of distinct topological structures.

We first explain how to achieve selective formation of either only skyrmions or only bubbles in a same specimen. Previous observations that bubbles are found in the presence of a small in-plane (IP) field [45-48] hint at the potential utility of an IP field for the selective realization of a skyrmion phase or a bubble phase. The bubble [Fig. 1(b)] can be stabilized by the IP field in a way that aligns the spins of the bubble boundary along the IP field. Given that skyrmions and bubbles are formed by pinching off stripe domains [45-48], applying the IP field in the direction of the stripe domain wall magnetization could result in the formation of the bubbles. Conversely, if one of the half circles of the bubble boundary [Fig. 1(b)] is reversed by reversing the IP field, the skyrmion structures could be achieved instead.

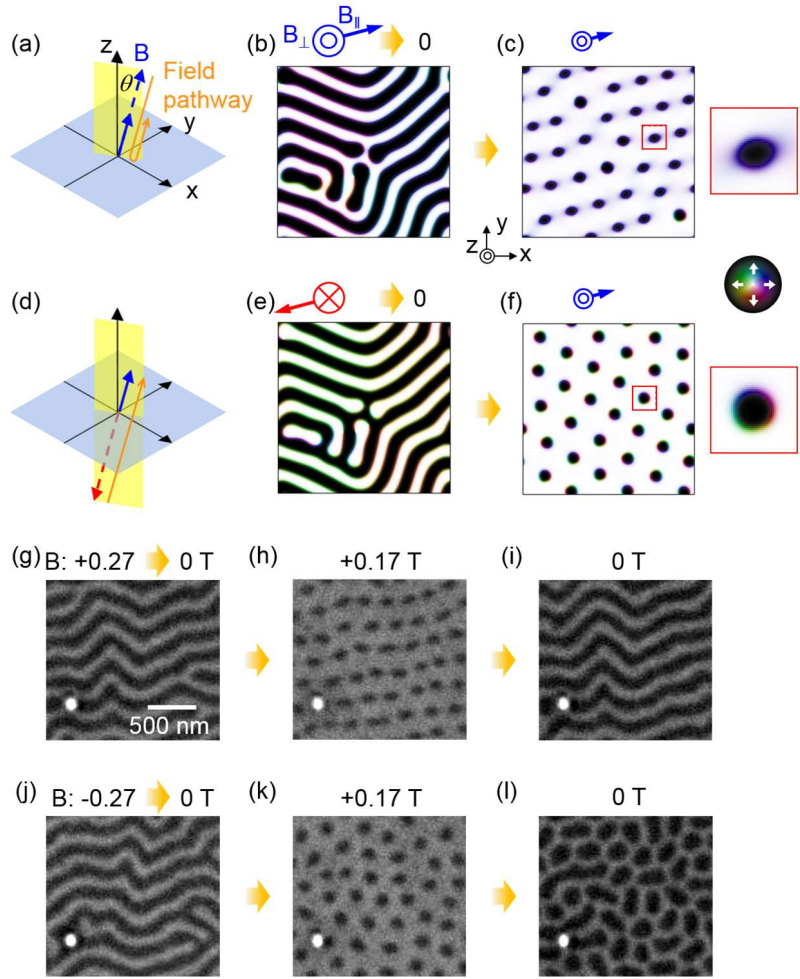

**FIG 2.** *Micromagnetic simulation and experimental results for the topology manipulation.* **(a-c)** Micromagnetic simulation for the bubble phase. The magnetic field configuration $B$ is illustrated in (a). The magnetization

images of the middle layer at zero field (b) and at the final magnetic field [+0.28 T, (c)]. The orange solid arrow in (a) represents the magnetic field pathway for the bubble phase. (d-f) Micromagnetic simulation for the skyrmion phase. The magnetic field configuration is illustrated in (d). The orange solid arrow in (d) represents the magnetic field pathway for the skyrmion phase. The magnetization at zero field (e) and the final magnetic field [+0.28 T, (f)] are shown. Zoomed images for a bubble (c) and a skyrmion (f) are also inserted. (g,h) MTXM images obtained for the pathway for the bubble generation depicted in (a). (i) Magnetic image at zero field when it is evolved from the bubble phase. (j,k) MTXM images obtained for the field pathway for the skyrmion generation depicted in (d). (l) Magnetic image at zero field when it is evolved from the skyrmion phase. The white spot in (g-l) is a structural defect, confirming that the bubble state and the skyrmion state are obtained in the same position.

We corroborate the above idea by micromagnetic simulations with magnetic parameters for the Fe/Gd multilayer film (see APPENDIX B). Fig. 2(a,d) schematically illustrate the designed pathways of the tilted magnetic field $B$ to realize the bubble phase and the skyrmion phase, respectively. Initially, stripe domains are prepared by applying a strong initialization field [dotted arrows in Fig. 2(a,d)] and then reducing the field to zero. Here the magnetic field $B$ is slightly tilted away from the z-axis ($\theta = 1°$) in order to provide a small IP field $B_\parallel$ and thus to line up the stripe domain wall magnetization along the IP field direction (ordered stripe phase). Fig. 2(b,e) present the magnetization of the middle layer of the ordered stripe phase obtained after the initializing field sequence. The in-plane magnetization of the domain wall is represented by the colours according to the depicted colour wheel, and it verifies that the ordered stripe phase is prepared as intended. Note that, in two different situations [Fig. 2(b,e)], the initial saturation fields are in the opposite direction but have the same magnitude so that the resultant stripe phases are identical upon the reversal of the film.

The tilted field $B$ is then increased until the stripe domains are pinched off. When the field is along the same direction as the saturation field and thus the IP field direction is not changed [Fig. (2a)], we obtain the bubble phase as expected [inset of Fig. 2(c)]. On the other hand, when the magnetic field is in the opposite direction to the saturation field and hence the IP component of the field changes its direction accordingly [Fig. 2(d)], we obtain the skyrmion phase of $|Q|=1$, as shown in the inset of Fig. 2(f). It is worth noting that completely different topological phases are obtained from basically the same initial stripe domain phase by following different IP field histories, suggesting the feasibility of the topology manipulation.

To experimentally confirm the above predictions, we investigate the Fe/Gd multilayered film at room temperature using the MTXM [49] at the Advanced Light Source (XM-1, BL6.1.2),

Lawrence Berkeley National Laboratory (see APPENDIX A). We follow the magnetic field pathways as described above. Fig. 2(g-i) and Fig. 2(j-l) correspond to the magnetic images representing the bubble case and the skyrmion case, respectively. For both cases, the ordered stripe phases are obtained at zero field [Fig. 2(g,j)].

When the field $B$ increases to +0.17 T, however, the final states are different from each other even at the same magnetic field [Fig. 2(h,k)]. Although isolated domains cover the entire film area for both cases, the shape of individual domains are slightly different. Fig. 2(k) shows circular domains of ~90 nm in diameter whereas Fig. 2(h) shows elongated domains. These results are in good agreement with the micromagnetic simulations [Fig. 2(c,f)], confirming that the bubble phase and the skyrmion phase are selectively realized in the single specimen. It should also be emphasized that the two distinct phases are obtained at the same magnetic field, enabling a fair comparison of the topological characteristics between the two topologically distinct phases. Furthermore, as an independent test, their topological differences are highlighted when the magnetic field comes back to zero as shown in Fig. 2(i,l). The bubble state [Fig. 2(h)] returns to an ordered stripe domain state [Fig. 2(i)] by connecting the bubbles in line. The skyrmion state [Fig. 2(k)], however, is topologically not allowed to merge, conserving their isolated features [Fig. 2(l)] even at zero field [28].

B. Experimental test of topological stability of skyrmions

Having established the two topologically inequivalent phases in the same specimen, we investigate their topological stabilities. We first study how differently the topologically distinct spin textures persist upon increasing the external magnetic field $B$ which shrinks those structures. As respectively shown in Fig. 3(a,b), the sizes of both bubbles and skyrmions reduce as the magnetic field increases, and they begin to disappear as the field is further increased. The normalized number of skyrmions (bubbles) $n_N$, remaining number divided by its initial number after the field application, and the radius of skyrmions (bubbles) as a function of the magnetic field $B$ are summarized in Fig. 3(c,d), respectively. Here the radius of elliptical bubbles is defined as the radius of a circle that has the same area of an ellipse. Two interesting aspects can be underlined. First, the number of skyrmions decreases much slower than that of bubbles. Second, upon increasing the applied field, a skyrmion survives to a much smaller size than a bubble. The size of skyrmion continuously decreases down to ~25 nm, whereas the bubble size decreases slowly to ~35 nm before the bubbles abruptly disappear. Considering the general correlation between the stability and the achievable minimum size, this result clearly shows superior stability of skyrmions to bubbles.

These results can be explained by the topological effect. The transformation of bubbles into a uniformly magnetized state is a continuous change of the spin structure that occurs within the same topological sector. However, the collapse of skyrmions into a uniform state requires a topological change, making the skyrmions persist at much stronger fields. Thus, the stronger field for the skyrmion annihilation corresponds to the extra field required to break the topology of the skyrmion structure. The topological change and corresponding extra field involve a rapid increase of the exchange-energy density at a Bloch point [11,15,35], which is a topologically-originated high energy barrier corresponding to the length scale of lattice constant, i.e., ultraviolet energy-cutoff between them, in continuum systems. In real discrete systems, the increased exchange energy is finite [8,14,16], but still large enough to guarantee a better stability for skyrmions than bubbles.

The topological stability is further supported by the significantly longer lifetimes of skyrmions than those of bubbles. The lifetime $\tau$ is directly related to the free energy barrier $\Delta F$, given as [40] $\tau(B) = \tau_{00} e^{\Delta F(B)/k_B T}$, where the prefactor $\tau_{00}$ is the characteristic time, $\Delta F = \Delta E - T\Delta S$, and the $k_B T$ is the thermal energy which is the product of the Boltzmann constant $k_B$ and temperature $T$. Here, $\Delta E$ and $\Delta S$ are the activation energy and entropy differences between the metastable state and the saddle point. In our cases where plenty of identical topological structures are present, the lifetime is represented by the decay of the number of spin textures with time.

To achieve a better temporal resolution, we apply a pulsed magnetic field $B_P$ as illustrated in Fig. 4(a), including typical magnetic images before and after the pulse field application. Either the skyrmion phase or the bubble phase is prepared under the initial field of +0.2 T, and the number of spin structures is counted ($N_{bef}$). A pulsed magnetic field $B_P$ is applied subsequently, and then the number of spin structures ($N_{aft}$) is counted again. It is confirmed that the initial field does not initiate the decay of spin textures, and the natural nucleation does not occur when the field comes back to the initial field level after the pulse field $B_P$. This guarantees that the decay events proceed only during the presence of the pulsed magnetic field $B_P$, thereby allowing us to estimate the decay rate with time by changing the duration $t$ of the $B_P$. The minimum duration which ensures a well-defined magnetic pulse shape is ~20 ms (see APPENDIX C). The decay rate $R_D$ is calculated by dividing the $N_{aft}$ by the $N_{bef}$, and the initial state for $N_{bef}$ is always initialized every time before the application of the $B_P$. For statistical significance, spin textures of $N_{bef} > 100$ in one area are analysed, and $R_D$ is averaged over 4 different areas in the film to obtain a single data point in Fig. (4).

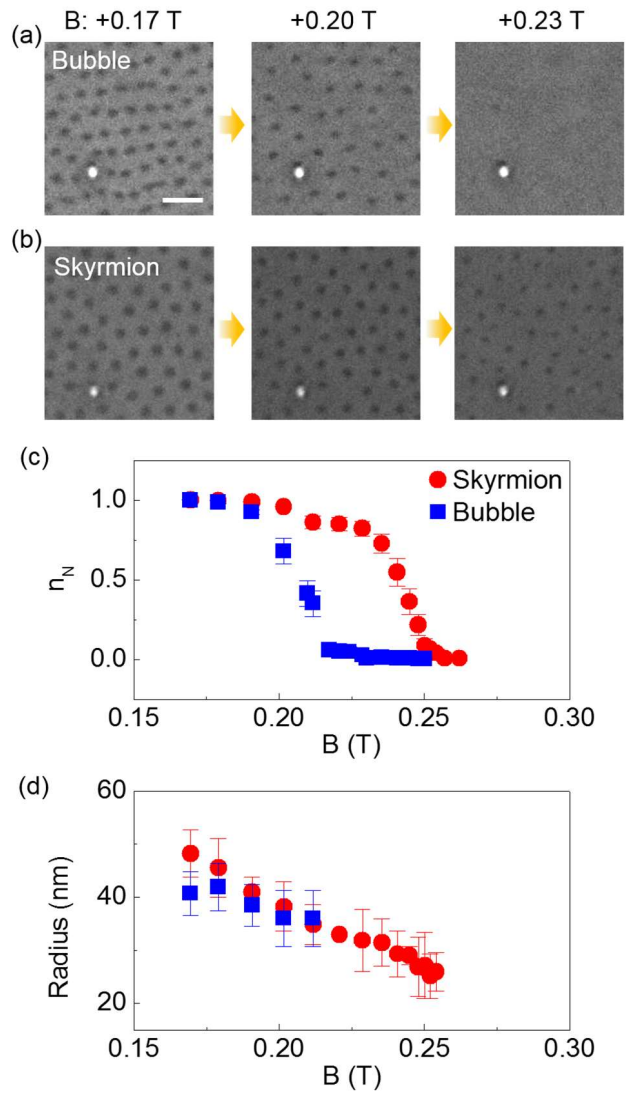

**Fig. 3.** *Magnetic field dependence of the bubble and the skyrmion phases.* **(a)** Evolution of the bubble phase with increasing fields. The scale bar is 500 nm. **(b)** Evolution of the skyrmion phase with increasing fields. **(c)** Annihilation of bubbles and skyrmions as a function of the magnetic field. The $n_N$ is the normalized number of spin structures which is divided by its initial number of spin textures. **(d)** Radius of bubbles and skyrmions as a function of the magnetic field.

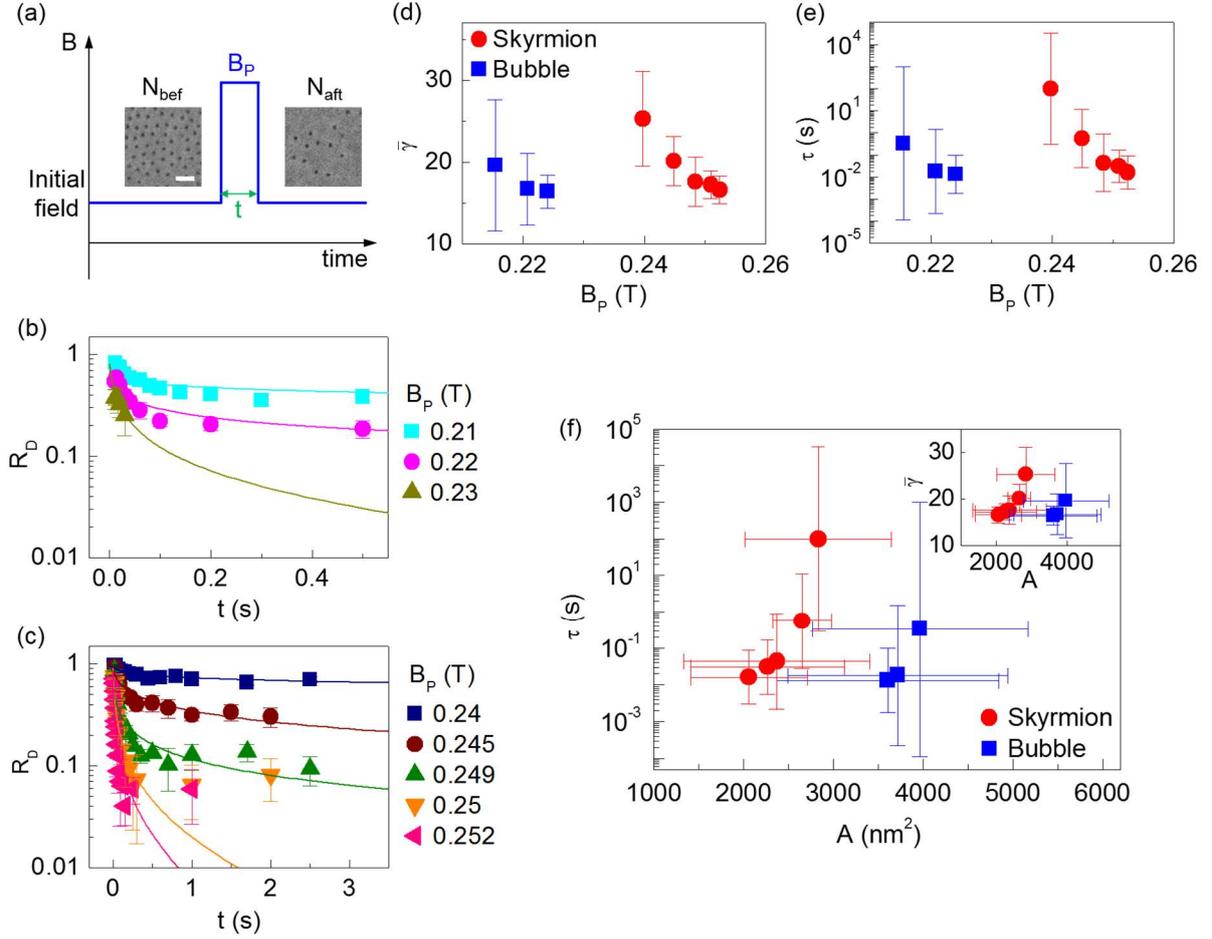

**Fig. 4.** *Time evolution of the decays of bubbles and skyrmions.* (a) Schematic of the experiment for the decay during the pulsed magnetic field $B_P$. The scale bar is 500 nm. (b,c) The decay rate $R_D$ of bubbles (b) and skyrmions (c) with respect to the duration $t$. Measured data points were fitted based on the exponential relationship and plotted in solid lines. (d,e) The reduced free energy barrier $\bar{\gamma}$ (d) and the lifetime $\tau$ (e) with respect to the field $B_P$. (f) The lifetime $\tau$ of bubbles and skyrmions as a function of their areas $A$. The inset shows the corresponding $\bar{\gamma}$ with respect to the area $A$. Red circles (blue squares) correspond the skyrmion (bubble). The error bars for $\bar{\gamma}$ and $\tau$ are determined by the standard deviation $\sigma$.

Fig. 4(b,c) summarize the decay rates $R_D$'s of bubbles and skyrmions under various magnetic fields as a function of time, respectively. The decay rate $R_D$ rapidly decreases with time in line with the previous work by Wild *et al.* [40], where an exponential time dependence of the skyrmion decaying process was observed. We fit the data (solid lines) by the exponential equation $R_D(t) = \int_0^\infty \frac{1}{\sqrt{2\pi\sigma^2}} exp\left[-\frac{(\gamma-\bar{\gamma})^2}{2\sigma^2}\right] exp\left[-\frac{t}{\tau_{00}exp(\gamma)}\right] d\gamma$, where the $\gamma$ ($\bar{\gamma}$) stands for the reduced free energy barrier $\Delta F/k_\mathrm{B}T$ (mean value of the $\gamma$), the $\sigma$ is the standard deviation of $\gamma$ and the $\tau_{00}$ is assumed to be $10^{-9}$ s. Gaussian distribution of $\gamma$ is assumed to account for the local variation of the magnetic properties in the sputtered film and consequently to improve the fit.

The obtained free energy barrier $\bar{\gamma}$ and the corresponding $\tau$, which is converted from $\bar{\gamma}$ using the Arrhenius law, are respectively shown in Fig. 4(d,e). Note that, while the $\bar{\gamma}$ is extracted under the assumption of the fixed $\tau_{00}$, the reconstructed $\tau$ does not contain any uncertainty which arises from the presumed $\tau_{00}$. Since the skyrmion decay occurs in a much stronger field range than the bubble case, the lifetimes could not be compared at the same magnetic fields. However, one can easily infer that the lifetime $\tau$ of skyrmions is much greater than that of the bubbles even at much stronger magnetic fields, in agreement with the result in Fig. 3(c).

The enhanced stability of skyrmions can also be highlighted when the $\tau$ is plotted as a function of the size (area, $A$) as shown in Fig. 4(f). Here, $A$ is calculated based on the data in Fig. 3(d), and linear extrapolation is adopted for the bubble area as the radius cannot be measured due to the fast annihilation of the bubbles. For both skyrmions and bubbles, it is noticed that the lifetime $\tau$ increases as the $A$ becomes larger, which is consistent with the general expectation that larger objects are more stable [1,34]. The most salient feature in Fig. 4(f) is that skyrmions (red circles) display much longer $\tau$ than bubbles (blue squares) despite their smaller sizes even at stronger fields, clearly indicating that the topological protection works for skyrmions, but not for bubbles. This also highlights the viability of magnetic skyrmions as topologically protected information carriers. The free energy barrier, $\bar{\gamma}$, with respect to the size is also presented in the inset of Fig. 4(f). The greater $\bar{\gamma}$ as well as the smaller sizes of skyrmions indicates that there is an additional factor associated with the topological change in skyrmion annihilation. Several works [37-41] have recently pointed out that entropic effect has an important role in the lifetime of skyrmions. In our present work, we are unable to separate the free energy barrier $\Delta F$ into the activation energy barrier $\Delta E$ and the entropy $\Delta S$. However, it does not alter our main conclusion; topological magnetic skyrmions have longer lifetimes and thus more stable than non-topological bubbles. The detailed separation between the energy barrier effect and the entropy effect would require temperature-dependent measurements [40], which is beyond the scope of this work.

III. CONCLUSIONS

To date, magnetic skyrmions have been the focus of extensive research because of the expectation for the exceptional topology-supported stability. This most important prerequisite of the skyrmion has always been taken for granted, but not been experimentally confirmed in comparison with a trivial spin structure. Also, as another great challenge to harness the topological nature of the materials, an efficient and versatile way to control the topology of the system has been elusive. In this work, we experimentally demonstrate that the topology of final states evolved from the same initial state can be controlled by choosing different pathways, enabling the manipulation of the topology in a given material system. This ability to control the topology of magnets would allow dynamic manipulation of various transport properties, e.g., Hall resistance which will be enhanced only in the skyrmion phase due to the topological Hall effect. By exploiting the revealed topology control, we demonstrate the existence of topological stability that has been often doubted in realistic discrete systems. We expect that our finding initiates further research efforts to tailor the topology of the system and enhance the topological energy barrier, which will facilitate the realization of ultrahigh density devices using topological spin structures.

**ACKNOWLEDGMENTS**

Works at the ALS were supported by U.S. Department of Energy (DE-AC02-05CH11231). SGJ, KJL and JIH acknowledges the support from the Future Materials Discovery Program through the NRF-2015M3D1A1070465. JIH and KSL also acknowledge the support from NRF-2017R1A2B4003139 and NRF-2019R1A2C2002996. SKK was supported by the start-up fund at the University of Missouri.

**APPENDIX A: SAMPLE PREPARATION AND IMAGING TECHNIQUE**

The amorphous Fe/Gd multilayer film was grown by magnetron sputtering at room temperature in a 3 mTorr Ar pressure at a base pressure of <3×10$^{-8}$ Torr. Fe and Gd layers were alternatively deposited on a 100 nm thick X-ray transparent Si$_3$N$_4$ membrane substrate and the process was repeated until the desired repetition number of 120 was achieved. The perpendicular magnetic contrast was measured by the X-ray magnetic circular dichroism (XMCD) at the Fe L$_3$ (708 eV) absorption edge in perpendicular geometry where the film surface is normal to the X-ray propagation direction. XMCD images were acquired using full-field magnetic transmission soft X-ray microscopy (MTXM) at the Advanced Light Source, Lawrence Berkeley National Laboratory (XM-1, beamline 6.1.2).

**APPENDIX B: MACROMAGNETIC SIMULATIONS**

Micromagnetic simulations were performed using the MuMax$^3$ code [50]. The model system was a 1,500×1,500×80 nm$^3$ plate with mesh sizes of 5×5×5 nm$^3$. For the simulations, typical magnetic parameters for the amorphous Fe/Gd ferrmimagnet [45], i.e., a saturation magnetization $M_S$ = 4×10$^5$ A/m, a uniaxial anisotropy $K_U$ = 4×10$^4$ J/m$^3$, and an exchange constant $A_{ex}$ = 5×10$^{-12}$ J/m, are used.

**APPENDIX C: PULSED MAGNETIC FIELD**

The minimum reliable duration of the pulsed magnetic field with a well-defined shape was confirmed by monitoring the current flowing through the coil for the magnetic field. The actual current pulse shape was measured by detecting the voltage across a test resistor of 0.5 Ω which is connected to the coil in series. As the resistance is sufficiently smaller than the resistance of the coil (8 Ω), the interference by the test resistor connection can be minimized. The detected voltage representing the current pulse profile for various durations (Supplementary Fig. S3) guarantees that the control of the duration is reliable.